%% file: template.tex
\documentclass[cameraready]{Interspeech}

\title{Adaptive Discovery of Interpretable Audio Attributes with Multimodal LLMs for Low-Resource Classification}

\author[correspondingauthor]{Kosuke}{Yoshimura}
\author[]{Hisashi}{Kashima}

\address{
    Kyoto University, Kyoto, Japan
}

\email{ykosuke@ml.ist.i.kyoto-u.ac.jp, kashima@i.kyoto-u.ac.jp}


\usepackage{comment}
\usepackage{algorithm}
\usepackage{algpseudocode}
\usepackage{color}
\usepackage{amsmath}
\usepackage{tabularx}
\usepackage{booktabs}
\usepackage{multirow}
\usepackage{amssymb}
\usepackage[table, dvipsnames]{xcolor} 
\definecolor{lightgray}{gray}{0.9} 

\algtext*{EndFor}
\algtext*{EndWhile}
\algtext*{EndIf}
\algtext*{EndFunction}

\begin{document}

\maketitle

\begin{abstract}
In predictive modeling for low-resource audio classification, extracting high-accuracy and interpretable attributes is critical. Particularly in high-reliability applications, interpretable audio attributes are indispensable. While human-driven attribute discovery is effective, its low throughput becomes a bottleneck.
We propose a method for adaptively discovering interpretable audio attributes using Multimodal Large Language Models (MLLMs). By replacing humans in the AdaFlock framework with MLLMs, our method achieves significantly faster attribute discovery. Our method dynamically identifies salient acoustic characteristics via prompting and constructs an attribute-based ensemble classifier.
Experimental results across various audio tasks demonstrate that our method outperforms direct MLLM prediction in the majority of evaluated cases. The entire training completes within 11 minutes, proving it a practical, adaptive solution that surpasses conventional human-reliant approaches.
\end{abstract}

\section{Introduction}





Extracting human-interpretable insights from complex acoustic signals remains a fundamental challenge in audio analysis. Even with the rise of large-scale generative models, interpretable attribute engineering remains essential for high-precision predictive modeling in low-resource scenarios. In such contexts, training models with massive numbers of parameters is often impractical; instead, lightweight models leveraging domain-specific audio attributes offer superior practicality and explainability. Additionally, in high-stakes domains, there is a strong demand for predictive reliability and explainability. In these fields, transparent models based on well-defined audio attributes are preferred over black-box inference. Consequently, in practical environments with strict constraints on computational resources and reliability, an approach centered on attribute engineering remains an extremely effective solution.

\begin{figure}[t]
  \centering
  \renewcommand{\arraystretch}{0.85}
  \includegraphics[width=\linewidth]{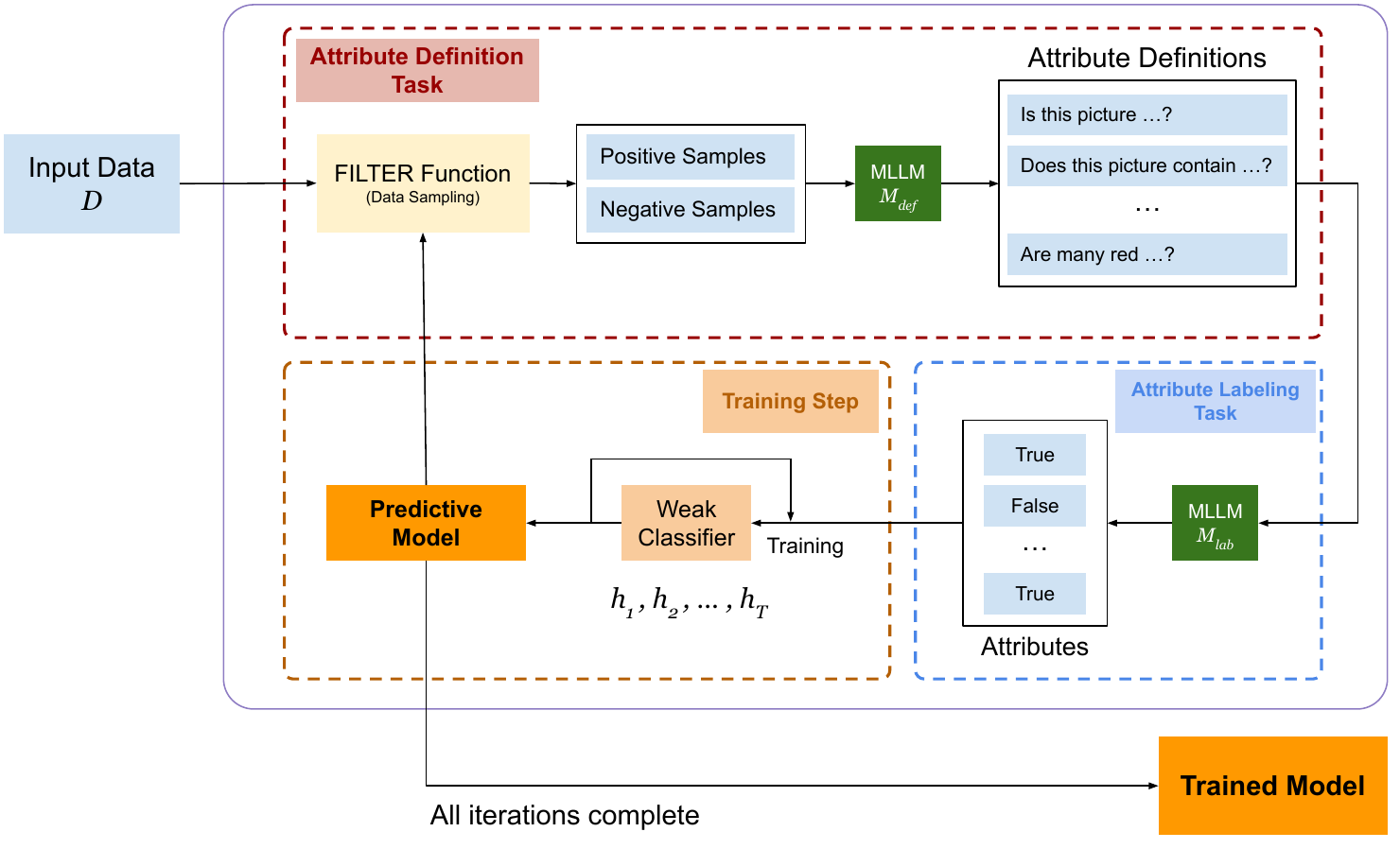}
  \caption{An iterative framework for MLLM-based attribute discovery and weak classifier training. Sampling weights are adaptively updated based on prediction results to discover attributes and refine model performance.}
  \label{fig:proposed_method}
\end{figure}

In this study, we investigate a method for efficiently defining and labeling audio attributes that achieve high predictive performance.  Traditionally, feature engineering for tabular data has often been limited to simple combinations of existing variables, making it difficult to extract highly creative features~\cite{traditional_method1, traditional_method2}. In contrast, crowdsourcing enables creative attribute discovery for unstructured data—such as audio, images, and text—by leveraging human cognitive abilities~\cite{flock, ada_flock}. However, traditional approaches in which humans handle the entire process from attribute ``definition'' to ``labeling'' face practical challenges, particularly requiring excessive lead times.
To address this issue, we propose a new method that uses Multimodal Large Language Models (MLLMs) to enable the adaptive discovery of interpretable audio attributes in a short time. Our approach follows an ``LLM-in-the-loop'' paradigm, where MLLMs serve as semantic oracles within a formal algorithmic framework, delegating high-level semantic judgments to the model while maintaining symbolic control structures. Inspired by AdaFlock \cite{ada_flock}, our approach replaces crowd workers with MLLMs to automate the discovery and labeling of audio data. Figure~\ref{fig:proposed_method} illustrates the overall architecture of our approach.
To demonstrate the utility of the proposed method, we conducted comparative experiments on predictive performance using four low-resource datasets of different modalities. Additionally, to compare with the lead time in crowdsourcing, we analyzed the training processing time.

This research makes three contributions: (1) an adaptive discovery method for interpretable audio attributes using MLLMs; (2) experimental results on four audio datasets showing that the attribute-based method outperforms direct MLLM inference under low-resource settings; and (3) a substantial reduction in the lead time for attribute discovery and labeling compared to crowdsourcing approaches.

\section{Related Work}

This section reviews attribute discovery, progressing from structured automation to human-centric and LLM-based methods. It also discusses the emerging LLM-in-the-loop paradigm, within which our proposed method is positioned.
Traditional automated approaches, such as Deep Feature Synthesis~\cite{traditional_method1, traditional_method2}, primarily target tabular data and rely on mathematical transformations within relational structures, making them difficult to apply to unstructured formats. To handle more diverse data, crowdsourcing frameworks like Flock~\cite{flock} and AdaFlock~\cite{ada_flock} leverage human cognitive abilities to extract creative features that complement model weaknesses. While effective, the high manual cost remains a significant barrier. Our method automates this human-centric process using MLLMs, achieving efficient attribute discovery across any data format while maintaining human-level interpretability.
Recently, LLM-based methods like CAAFE \cite{caafe} and iterative optimization scripts \cite{llm_feature_creation} have introduced semantic reasoning into feature engineering. However, these state-of-the-art techniques still primarily focus on tabular metadata and lack the capability to directly analyze raw multimodal data. By automating the adaptive reasoning traditionally performed in crowdsourcing with MLLMs, our proposed method enables versatile attribute discovery that transcends constraints imposed by traditional data formats. Our approach aligns with the LLM-in-the-loop paradigm seen in query processing~\cite{chengbinding, glenn2024blendsql} and clustering~\cite{zhang-etal-2023-clusterllm, diaz2026kLLMmeans}, where LLMs are integrated as internal modules for semantic judgment within an explicit algorithmic structure.

\section{Problem Setup}
Let $\mathcal{D} = \{(x_i, y_i)\}_{i=1}^{N}$ be a labeled training dataset, where each $x_i \in \mathcal{X}^{\text{audio}}$ is a variable-length audio signal (e.g., speech or other acoustic events) and $y_i \in \{0,1\}$ is the corresponding label.
We consider a low-resource audio classification setting in which the number of labeled training samples $N$ is small (e.g., on the order of hundreds). Our objective is to learn a classifier $H: \mathcal{X}^{\text{audio}} \rightarrow \{0,1\}$ that generalizes to unseen audio inputs.

In such low-resource scenarios, directly training or fine-tuning large-scale end-to-end models is often computationally expensive and prone to overfitting due to data scarcity. In addition to $\mathcal{D}$, we assume access to pretrained foundation models capable of processing audio signals. Rather than fine-tuning these models, we use them to define a set of semantically meaningful binary attribute definitions and to assign attribute labels to each audio signal. The learning problem is therefore to construct a data-efficient classifier on top of these foundation-model-derived attributes under limited labeled data.

\section{Proposed Method}
\input{algo}
We propose a three-stage framework using two MLLMs, $\mathcal{M}_{def}$ and $\mathcal{M}_{lab}$, to discover data-adaptive attributes: (1) Attribute Definition: $\mathcal{M}_{def}$ defines new concepts by prioritizing samples where the current model fails. (2) Attribute Labeling:  $\mathcal{M}_{lab}$ performs labeling for each data point based on the defined attribute to construct the dataset for prediction. (3) Weak Classifier Training: A boosting model is trained on these discovered attributes. Repeating this cycle yields the final attribute set and model, as detailed in Figure~\ref{fig:proposed_method} and Algorithm~\ref{alg:proposed_method}.

\subsection{Sampling and Attribute Definition via \texorpdfstring{$\mathcal{M}_{def}$}{M\_def}}
$\mathcal{M}_{def}$ functions as an autonomous discovery engine, leveraging open-vocabulary knowledge to identify latent discriminative patterns. Crucially, $\mathcal{M}_{def}$ is provided only with the grouped samples (Group A and Group B) without any explicit class labels or domain-specific metadata. This ensures that the attribute discovery process is driven purely by the acoustic contrast between the two sets, allowing the model to induce salient attributes in a bottom-up, data-driven manner. To initiate this process, $q$ positive and $q$ negative examples are grouped and presented to $\mathcal{M}_{def}$. The model is tasked with generating $k$ descriptions to distinguish between the two groups, using a prompt such as: ``Create $k$ attributes that describe the differences between Group A and Group B.'' The prompt ensures each description is a yes/no question, a binary format that aligns with weak classifiers like decision trees, which partition the attribute space via discrete attributes. Furthermore, this approach ensures high interpretability, as the final ensemble decision can be traced back to a series of intuitive linguistic queries discovered without any prior label bias.

To implement this definition process effectively, the set of examples presented to $\mathcal{M}_{def}$ is constructed using weighted sampling based on the current model's predictive performance. This sampling is managed by a ``FILTER'' function that implements a rejection-sampling-based strategy, as detailed in Algorithm \ref{alg:proposed_method}.
Specifically, for an example $x_i$ randomly drawn from the dataset, an acceptance test is performed where $x_i$ is accepted with a probability proportional to its weight  $w_i' \in [0, 1]$. Only accepted examples are included in the sample set. 
Following the learning results of the boosting process described later, these weights $w_i'$ are updated so that examples incorrectly predicted by the current model receive higher weights. Consequently, hard examples are prioritized for inclusion in the presented samples. This error-centric sampling ensures that $\mathcal{M}_{def}$ focuses on the current model's blind spots, synthesizing attributes that specifically refine the decision boundary where the model is most prone to error. This process is repeated for each attribute definition until the sample set contains exactly $q$ positive and $q$ negative examples.

\subsection{Attribute Labeling via \texorpdfstring{$\mathcal{M}_{lab}$}{M\_lab}}
We present $k$ yes/no questions discovered during the attribute definition phase for all training instances at each iteration, and have $\mathcal{M}_{lab}$ determine their truth values. To reduce computational cost, instead of making $N \cdot k$ separate queries, we present all $k$ questions simultaneously. This limits the total number of MLLM queries to $N$ per attribute labeling session.

\subsection{Training Weak Classifiers via Boosting}
We perform adaptive learning by applying the AdaBoost framework using attributes sequentially discovered by MLLMs. At each iteration $t = 1, \dots, T$, we first calculate the edge $\gamma_t$ of a weak classifier $h_t$—trained based on the attributes $F_t$ defined by MLLMs—under the current weight distribution $w^t$ as $\gamma_t = (\sum_{i=1}^N \mathbf{1}[y_i = h_t(x_i)] w_i^t)/(\sum_{i=1}^N w_i^t) - 0.5$, where $\mathbf{1}[\cdot]$ denotes the indicator function.
Using this $\gamma_t$, the weight $\alpha_t$ representing the confidence of the weak classifier $h_t$ is determined as $\alpha_t = 0.5 \ln ( (0.5 + \gamma_t) / (0.5 - \gamma_t) )$. The obtained $\alpha_t$ is used to update the provisional ensemble classifier $H_t(x) = H_{t-1}(x) + \alpha_t h_t(x)$. Accordingly, the weight of each instance $w_i^{(t+1)}$ is updated based on the logistic loss as $w_i^{(t+1)} = (1 + \exp(y_i H_t(x_i)))^{-1}$.
The weights calculated here are normalized to sum to 1 in the next step and, via a FILTER function, function as a probability distribution to preferentially sample ``hard instances'' that the current model misclassifies. Finally, the ensemble classifier $H_T(x) = \text{sign} ( \sum_{t=1}^T \alpha_t h_t(x) )$, which performs the final prediction, is constructed by taking the sign of the weighted sum of the weak classifiers obtained over all $T$ iterations. 

\subsection{Inference}
Since the proposed method defines new attributes, they must be assigned to each instance prior to inference. Therefore, for unseen instances, attribute labeling using $\mathcal{M}_{lab}$ is performed first, followed by inference via the ensemble classifier $H_T$. Although this process introduces MLLM latency during inference, the computational overhead is highly parallelizable as each instance can be labeled independently. Furthermore, query efficiency can be optimized by batching multiple instances, mirroring the strategy used during training to ensure practical feasibility.

\section{Experiments}
In this section, we evaluate the proposed method's performance in comparison with MLLMs and feature-based baselines. To ensure the reliability of our findings, we report the mean performance over ten independent trials for all stochastic processes. Our implementation is available at [Anonymous Link]\footnote{The code will be released upon paper acceptance.}.

\subsection{Datasets}
We evaluated our method using \textbf{CREMA-D} \cite{cremad}, \textbf{RAVDESS} \cite{ravdess}, \textbf{Coswara} \cite{coswara}, and \textbf{ESC-50} \cite{esc50}, each adapted for binary classification with balanced training and test sets ($n=100$ per class). For CREMA-D (91 actors) and RAVDESS (24 actors), \textit{Happy} was designated as Positive, while Negative instances combined \textit{Sad}, \textit{Angry}, \textit{Disgust}, and \textit{Fear} (CREMA-D) or \textit{Fearful} (RAVDESS). For Coswara, we selected \textit{cough-heavy.wav} samples with quality scores of 1--2, labeling \textit{positive\_mild/moderate/asymp} as Positive and \textit{healthy} as Negative, with gender-balanced sets (50 males/50 females per class). For ESC-50, we defined two classes by grouping ten categories: \textit{gaseous-fluid} (wind, vacuum cleaner, airplane, helicopter, breathing) and \textit{liquid-fluid} (rain, sea waves, pouring water, washing machine, drinking sipping), using stratified sampling to maintain category distribution.

\subsection{Experimental Setup}
\textbf{Models:} We employ three audio-native models as backbones for both our proposed method and the MLLM prediction baseline: \textit{GPT-audio-mini}~\cite{gpt_audio_mini} (\texttt{gpt-audio-m}) and \textit{GPT-audio}~\cite{gpt_audio} (\texttt{gpt-audio}) from OpenAI, and \textit{Gemini-3-flash-preview}~\cite{gemini} (\texttt{gemini-3-f-p}) from Google.
\textbf{Baselines:} The proposed method is compared against two baselines: (1) \textbf{LR}: Logistic Regression using CLAP~\cite{clap} features with an L-BFGS solver (via scikit-learn~\cite{sklearn}), and (2) \textbf{MLLM prediction}: direct zero-shot prediction by the MLLMs.
\textbf{Evaluation Protocol:} To ensure reliability against the stochastic nature of MLLM generation, we report the mean accuracy over ten trials with the sampling temperature set to $1.0$. For the LR baseline, a single representative score is reported as both the CLAP feature extraction and the L-BFGS solver (with a fixed initialization) are deterministic.
\textbf{Hyperparameters:} For our method, we set $q=10$ (positive/negative examples), $T=10$ (total iterations), and $k=10$ (attributes per iteration) for all datasets. A decision stump~\cite{decision_stump} is utilized as the weak classifier.

\subsection{Performance Comparison}

\begin{table}[t]
\centering
\small
\setlength{\tabcolsep}{3pt}
\renewcommand{\arraystretch}{0.85}
\caption{Comparison of the proposed method against two baselines (MLLM prediction and LR). Values represent mean accuracy $\pm$ standard deviation (\%), except for the LR baseline which is reported as a single deterministic score. $\Delta$ indicates the absolute improvement of Ours over the MLLM Prediction baseline. \textbf{Bold} indicates the best performance for each dataset.}
\label{tab:accuracy}
\addtolength{\tabcolsep}{-3pt}
\begin{tabular*}{\columnwidth}{@{\extracolsep{\fill}}lcccc}
\toprule
& \multicolumn{2}{c}{\textbf{Baselines}} & \multicolumn{2}{c}{\textbf{Proposed}} \\ \cmidrule(lr){2-3} \cmidrule(lr){4-5}
\textbf{Dataset} & \textbf{LR} & \textbf{MLLM Pred.} & \textbf{Ours} & \textbf{$\Delta$} \\
\midrule
CREMA-D & 70.00 & 69.00 $\pm$ 2.32 & \textbf{72.45 $\pm$ 6.10} & +3.45 \\
RAVDESS & 54.00 & 66.60 $\pm$ 1.82 & \textbf{68.55 $\pm$ 2.57} & +1.95 \\
Coswara & \textbf{67.50} & 48.10 $\pm$ 2.00  & 55.70 $\pm$ 2.20 & +7.60 \\
ESC-50  & \textbf{94.00} & 88.35 $\pm$ 0.88  & 87.15 $\pm$ 1.49 & -1.20 \\
\bottomrule
\end{tabular*}
\end{table}

Table~\ref{tab:accuracy} presents the comparative results of our evaluation. To ensure a fair comparison and isolate the impact of our framework, we employed \texttt{gemini-3-f-p} as the core model for both the MLLM prediction baseline and the proposed method.
The proposed method consistently outperformed the MLLM prediction on three of the four datasets, showing accuracy gains in Coswara (+7.60\%) and CREMA-D (+3.45\%). These results underscore the effectiveness of our approach in enhancing the inference capabilities of the MLLM.

However, a comparison with the LR baseline reveals a more nuanced performance profile. Our method demonstrates a clear advantage over LR in emotion recognition tasks, specifically for CREMA-D (72.45\% vs. 70.00\%) and RAVDESS (68.55\% vs. 54.00\%). In contrast, the LR baseline remains superiority in other domains; it achieved 94.00\% accuracy on the environmental sound task (ESC-50) and 67.50\% on the medical audio dataset (Coswara), both exceeding the proposed method’s performance.
Overall, the proposed method demonstrates strong effectiveness in low-resource speech 
classification tasks where discriminative information is largely derived from 
semantic and conceptual representations. Conversely, in tasks dominated by 
low-level acoustic statistics, continuous acoustic embeddings appear to provide 
stronger discriminative power.

\subsection{Qualitative Analysis: Discovered Attributes}
\begin{table}[t]
    \centering
    \footnotesize
    \setlength{\tabcolsep}{3pt}
    \renewcommand{\arraystretch}{0.85}
    \setlength{\tabcolsep}{3pt}
    \caption{Top 3 attributes discovered by \texttt{gemini-3-f-p} for each dataset, listed in descending order of importance.}
\label{tab:discovered_features}
    \begin{tabularx}{\columnwidth}{lX}
        \toprule
        \textbf{Dataset} & \textbf{Attributes} \\
        \midrule
        CREMA-D & Is the speaker's tone generally positive or upbeat?;
Does the speaker sound relaxed rather than intense or secretive?;
Does the speaker's voice sound cheerful?;\\
        \addlinespace
        RAVDESS & Does the speaker's tone suggest more of a positive-neutral emotion, as opposed to a negative-neutral emotion?;
Does the speaker sound excited?;
Does the speaker's voice exhibit signs of distress or agitation?;\\
        \addlinespace
        Coswara & Is the cough followed by a distinct audible breath intake?;
Does the cough appear to be prolonged or linger after it starts?;
Does the frequency of coughing increase over the duration of the sample?;\\
        \addlinespace
        ESC-50  & Does the environment sound windy or breezy?;
Is there a presence of moving water in the background?;
Does the sound contain high-pitched elements like splashes or drips?;\\
        \bottomrule
    \end{tabularx}
\end{table}

Table~\ref{tab:discovered_features} summarizes attributes extracted by \texttt{gemini-3-f-p}, where importance is defined by the reliability score $\alpha_t$. Despite no label access, the MLLM identifies attributes with strong semantic alignment to ground-truth concepts. For emotional speech (CREMA-D, RAVDESS), it autonomously captured valence (e.g., ``upbeat'', ``cheerful'') and arousal (e.g., ``excited'', ``distress or agitation'') indicators. In Coswara, the model extracted clinically relevant markers such as ``audible breath intake'' and temporal changes in cough frequency. For ESC-50, attributes characterized physical textures like ``windy'' or ``moving water.'' These results validate that our method maps raw audio into a structured, interpretable linguistic space that justifies ensemble decisions.

\subsection{Analysis of MLLM Variations in Attribute Definition}
\begin{table}[t]
\centering
\footnotesize
\setlength{\tabcolsep}{3pt}
\renewcommand{\arraystretch}{0.85}
\caption{Impact of Different Attribute Definition Models ($\mathcal{M}_{def}$) on Accuracy. Values represent mean accuracy $\pm$ standard deviation (\%). $\Delta_{max}$ denotes the difference between the highest and lowest mean accuracy among $\mathcal{M}_{def}$ models. \textbf{Bold} values indicate the best performance for each dataset.}
\label{tab:m_def_impact}
\addtolength{\tabcolsep}{-3pt}
\begin{tabular*}{\columnwidth}{@{\extracolsep{\fill}}lcccc}
\toprule
& \multicolumn{3}{c}{\textbf{Attribute Definition Model ($\mathcal{M}_{def}$)}} & \\
\cmidrule(lr){2-4}
\textbf{Dataset} & \textbf{gpt-audio-m} & \textbf{gpt-audio} & \textbf{gemini-3-f-p} & \textbf{$\Delta_{max}$} \\
\midrule
CREMA-D & 73.55 $\pm$ 4.05 & \textbf{75.30 $\pm$ 3.94} & 72.45 $\pm$ 6.10  & 2.88 \\
RAVDESS & \textbf{68.65 $\pm$ 3.81} & 66.90 $\pm$ 1.63 & 68.55 $\pm$ 2.57  & 1.75 \\
Coswara & 54.20 $\pm$ 1.84 & 53.35 $\pm$ 3.13 & \textbf{55.70 $\pm$ 2.20}  & 2.35 \\
ESC-50  & 87.25 $\pm$ 1.99 & \textbf{88.10 $\pm$ 2.07} & 87.15 $\pm$ 1.49  & 0.95 \\
\bottomrule
\end{tabular*}
\end{table}
Table~\ref{tab:m_def_impact} illustrates the impact of different MLLMs used for attribute definition ($\mathcal{M}_{def}$), with $\mathcal{M}_{lab}$ fixed to \texttt{gemini-3-f-p}. No single model consistently dominated across datasets, though \texttt{gpt-audio} achieved the highest accuracy in emotion recognition (75.30\% on CREMA-D) and environmental sound classification (88.10\% on ESC-50). \texttt{gemini-3-f-p} and \texttt{gpt-audio-m} also showed competitive performance, securing the top results for Coswara and RAVDESS, respectively. These results suggest that while model capacity influences attribute quality, the optimal $\mathcal{M}_{def}$ depends on the acoustic characteristics of the task. Notably, the performance gap remains minimal across all datasets, with a maximum difference ($\Delta_{max}$) of only 2.88\%. This consistency demonstrates that our framework is highly robust to the choice of the definition model, provided it possesses sufficient multimodal reasoning capabilities to translate audio attributes into descriptive text.

\subsection{Training Time Analysis}
\begin{table}[t]
\centering
\footnotesize
\setlength{\tabcolsep}{4pt}
\renewcommand{\arraystretch}{0.85}
\caption{Training time [min] (mean $\pm$ standard deviation) per dataset across different $\mathcal{M}_{def}$. $\mathcal{M}_{lab}$ is fixed to \texttt{gemini-3-f-p}.}
\label{tab:processing_time}
\setlength{\tabcolsep}{3pt} 
\begin{tabular*}{\columnwidth}{@{\extracolsep{\fill}}lccc}
\toprule
& \multicolumn{3}{c}{\textbf{Attribute Definition Model ($\mathcal{M}_{def}$)}} \\ \cmidrule(lr){2-4}
\textbf{Dataset} & \textbf{gpt-audio} & \textbf{gpt-audio-m} & \textbf{Gemini-3-f-p} \\
\midrule
CREMA-D & 7.89 $\pm$ 1.00 & 7.72 $\pm$ 1.11 & 8.57 $\pm$ 0.81 \\
RAVDESS & 9.87 $\pm$ 1.56 & 9.74 $\pm$  1.81 & 8.84 $\pm$ 1.35 \\
Coswara & 10.53 $\pm$ 1.54 & 10.30 $\pm$ 1.78 & 8.43 $\pm$ 0.84 \\
ESC-50 & 9.63 $\pm$ 1.02 & 9.50 $\pm$ 0.68 & 9.11 $\pm$ 0.92 \\

\bottomrule
\end{tabular*}
\end{table}
The primary advantage of the proposed method lies in its ability to drastically reduce the time required to complete training compared to conventional human-in-the-loop approaches. As Table \ref{tab:processing_time} shows, the maximum mean processing time was approximately 10.53 minutes for the Coswara dataset (using \texttt{gpt-audio} as $\mathcal{M}_{def}$), with other configurations requiring less time (7.72--10.53 minutes).
Unlike human-in-the-loop studies slowed by recruitment and response delays, our purely computational approach eliminates such overheads. This efficiency enables rapid trial-and-error within practical timeframes.

\section{Conclusion}
This paper proposed an adaptive discovery method for interpretable audio attributes using MLLMs. By integrating MLLMs into the definition and labeling processes, our approach automatically discovers discriminative attributes for low-resource tasks. Evaluations across four datasets showed that our method outperformed MLLM prediction in three cases. Notably, in emotion recognition tasks (CREMA-D and RAVDESS), it also surpassed the LR baseline using CLAP embeddings. Analysis of MLLM variations demonstrated the framework's robustness, with final accuracy showing minimal divergence across different attribute definition models. Finally, the entire process took less than 11 minutes, offering a practical and interpretable alternative to human-reliant attribute engineering.
\newpage
\section{Generative AI Use Disclosure}
We employed Generative AI tools (ChatGPT and Gemini) exclusively for the purpose of refining the manuscript’s grammar and clarity.
\bibliographystyle{IEEEtran}
\bibliography{mybib}

\end{document}

%% file: algo.tex
\begin{algorithm}[t]
\caption{Adaptive Attribute Discovery with MLLMs.}
\label{alg:proposed_method}
\small
\begin{algorithmic}[1]
\Require MLLMs $\mathcal{M}_{def}$ (for definition) and $\mathcal{M}_{lab}$ (for labeling), training data set $D=\{(x_i,y_i)\}_{i=1}^N$, termination parameters $\delta_t$ and $\epsilon$, number of samples per label per iteration $q$, number of attribute definitions per iteration $k$, number of iterations $T$
\Ensure Ensemble classifier $H_T$

\State For each $i=1,\ldots,N$, initialize weights $w_i^{1} \gets 1/N$
\State Initialize ensemble classifier $H_0(x) \gets 0$
\For{$t = 1$ to $T$}
\State $S_F^+ \gets \emptyset$, $S_F^- \gets \emptyset$, $r \gets 0$
\While{$|S_F^+| < q$ \textbf{or} $|S_F^-| < q$}
  \State Sample $(x,y,r)$ using $\textsc{FILTER}(D, w^{t}, r, \delta_t, \epsilon)$
  \If{$x = \bot$} \textbf{continue}
\EndIf
  \If{$y=+1$ \textbf{and} $|S_F^+| < q$}
    \State $S_F^+ \gets S_F^+ \cup \{(x,y)\}$
  \ElsIf{$y=-1$ \textbf{and} $|S_F^-| < q$}
    \State $S_F^- \gets S_F^- \cup \{(x,y)\}$
  \EndIf
\EndWhile

    \State $S_F \gets S_F^+ \cup S_F^-$
    \State Obtain attribute definitions $F_t=\{f_t^1,\ldots,f_t^k\}$ from $S_F$ using $\mathcal{M}_{def}$
    \State Label $D$ based on $F_t$ using $\mathcal{M}_{lab}$
    \State Train weak learner $h_t$ based on $F_t$

    \State $\gamma_t = (\sum_{i=1}^N \mathbf{1}[y_i = h_t(x_i)] w_i^t)/(\sum_{i=1}^N w_i^t) - 0.5$
    \State $\displaystyle
    \alpha_t \gets 0.5 \ln ( (0.5 + \gamma_t) / (0.5 - \gamma_t) )$
    \State $H_t(x) \gets H_{t-1}(x) + \alpha_t h_t(x)$
    \State For all $i$, $w_i^{(t+1)} \gets (1+\exp(y_i H_t(x_i)))^{-1}$
    \State Normalize $w^{t+1} \gets w^{t+1} / \sum_i w_i^{t+1}$
\EndFor

\State \Return $H_T(x) = \mathrm{sign} \left(\sum_{t=1}^T \alpha_t h_t(x)\right)$

\State
\Function{FILTER}{$D, w^{t}, r, \delta_t, \epsilon$}
  \State $r \gets r + 1$ \Comment{Number of calls in step $t$}
  \State $\delta'_t \gets \frac{\delta_t}{r(r+1)}$
    \State $(D',w') \gets$ Random permutation of $(D, w^t)$
    \For{$i = 0; i < \min(N, \frac{2}{\epsilon}\ln(1/\delta'_t)); i = i + 1$}
        \State $(x,y) \gets i$-th element of $D'$
        \State \textbf{return} $(x, y, r)$ with probability $w'_i$
    \EndFor
  \State \Return $(\bot, \bot, r)$
\EndFunction
\end{algorithmic}
\end{algorithm}

%% file: mybib.bib
@INPROCEEDINGS{traditional_method1,
  author={Kanter, James Max and Veeramachaneni, Kalyan},
  booktitle={DSAA 2015}, 
  title={Deep feature synthesis: Towards automating data science endeavors}, 
  year={2015},
}

@InProceedings{traditional_method2,
author="Horn, Franziska
and Pack, Robert
and Rieger, Michael",
title="The autofeat Python Library for Automated Feature Engineering and Selection",
booktitle="ECML PKDD 2019",
year="2020",
}

@inproceedings{flock,
author = {Cheng, Justin and Bernstein, Michael S.},
title = {Flock: Hybrid Crowd-Machine Learning Classifiers},
year = {2015},
booktitle = {CSCW 2015},
}

@inproceedings{ada_flock,
  title={AdaFlock: Adaptive Feature Discovery for Human-in-the-loop Predictive Modeling},
  author={Takahama, Ryusuke and Baba, Yukino and Shimizu, Nobuyuki and Fujita, Sumio and Kashima, Hisashi},
  booktitle={AAAI 2018},
  year={2018}
}

@inproceedings{caafe,
author = {Hollmann, Noah and M\"{u}ller, Samuel and Hutter, Frank},
title = {Large language models for automated data science: introducing CAAFE for context-aware automated feature engineering},
year = {2023},
booktitle={NeurIPS 2023},
}

@inproceedings{llm_feature_creation,
author = {Nam, Jaehyun and Kim, Kyuyoung and Oh, Seunghyuk and Tack, Jihoon and Kim, Jaehyung and Shin, Jinwoo},
title = {Optimized feature generation for tabular data via LLMs with decision tree reasoning},
year = {2024},
booktitle={NeurIPS 2024},
}

@ARTICLE{cremad,
  author={Cao, Houwei and Cooper, David G. and Keutmann, Michael K. and Gur, Ruben C. and Nenkova, Ani and Verma, Ragini},
  journal={IEEE Trans. Affect. Comput.},
  title={CREMA-D: Crowd-Sourced Emotional Multimodal Actors Dataset}, 
  year={2014},
}

@misc{ravdess,
  author       = {Livingstone, Steven R. and
                  Russo, Frank A.},
  title        = {The Ryerson Audio-Visual Database of Emotional
                   Speech and Song (RAVDESS)
                  },
  year         = 2018,
  publisher    = {Zenodo},
  version      = {1.0.0},
  url          = {https://doi.org/10.5281/zenodo.1188976},
}

@misc{coswara,
  author       = {Debarpan Bhattacharya and
                  Neeraj Kumar Sharma and
                  Debottam Dutta and
                  Srikanth Raj Chetupalli and
                  Pravin Mote and
                  Sriram Ganapathy and
                  Chandrakiran C and
                  Sahiti Nori and
                  Suhail K K and
                  Sadhana Gonuguntla and
                  Murali Alagesan},
  title        = {Coswara: A respiratory sounds and symptoms dataset
                   for remote screening of SARS-CoV-2 infection
                  },
  year         = 2022,
  version      = {v.1.0},
  url          = {https://doi.org/10.5281/zenodo.7188627},
}

@inproceedings{esc50,
  title = {{ESC}: {Dataset} for {Environmental Sound Classification}},
  author = {Piczak, Karol J.},
  booktitle = {MM 2015},
  year = {2015}
}

@INPROCEEDINGS{clap,
  author={Wu, Yusong and Chen, Ke and Zhang, Tianyu and Hui, Yuchen and Berg-Kirkpatrick, Taylor and Dubnov, Shlomo},
  booktitle={ICASSP 2023}, 
  title={Large-Scale Contrastive Language-Audio Pretraining with Feature Fusion and Keyword-to-Caption Augmentation}, 
  year={2023},
  }

@article{decision_stump,
title = {A Decision-Theoretic Generalization of On-Line Learning and an Application to Boosting},
journal = {Journal of Computer and System Sciences},
year = {1997},
author = {Yoav Freund and Robert E Schapire},
}

@article{sklearn,
  author  = {Fabian Pedregosa and Ga{{\"e}}l Varoquaux and Alexandre Gramfort and Vincent Michel and Bertrand Thirion and Olivier Grisel and Mathieu Blondel and Peter Prettenhofer and Ron Weiss and Vincent Dubourg and Jake Vanderplas and Alexandre Passos and David Cournapeau and Matthieu Brucher and Matthieu Perrot and {{\'E}}douard Duchesnay},
  title   = {Scikit-learn: Machine Learning in Python},
  journal = {Journal of Machine Learning Research},
  year    = {2011},
}

@misc{gpt_audio,
  author       = {OpenAI},
  title        = {{GPT}-audio},
  year         = {2024},
  howpublished = {\url{https://developers.openai.com/api/docs/models/gpt-audio}},
  note         = {Accessed: 2026-03-04}
}

@misc{gpt_audio_mini,
  author       = {OpenAI},
  title        = {{GPT}-audio-mini},
  year         = {2024},
  howpublished = {\url{https://developers.openai.com/api/docs/models/gpt-audio-mini}},
  note         = {Accessed: 2026-03-04}
}

@misc{gemini,
  author       = {Google},
  title        = {Gemini 3},
  year         = {2024},
  howpublished = {\url{https://ai.google.dev/gemini-api/docs/models}},
  note         = {Accessed: 2026-03-04}
}

@inproceedings{chengbinding,
  title     = {Binding Language Models in Symbolic Languages},
  author    = {Cheng, Zhoujun and Xie, Tianbao and Shi, Peng and Li, Chengzu and 
               Nadkarni, Rahul and Hu, Yushi and Xiong, Caiming and 
               Radev, Dragomir and Ostendorf, Mari and Zettlemoyer, Luke},
  booktitle = {ICLR 2023},
  year      = {2023}
}

@inproceedings{glenn2024blendsql,
  title     = {Blend{SQL}: A Scalable Dialect for Unifying Hybrid Question Answering in Relational Algebra},
  author    = {Glenn, Parker and Dakle, Parag and Wang, Liang and Raghavan, Preethi},
  booktitle = {ACL 2024},
  year      = {2024}
}

@inproceedings{zhang-etal-2023-clusterllm,
  title     = {ClusterLLM: Large Language Models as a Guide for Text Clustering},
  author    = {Zhang, Yuwei and Wang, Zihan and Shang, Jingbo},
  booktitle = {EMNLP 2023},
  year      = {2023}
}

@inproceedings{diaz2026kLLMmeans,
  title     = {Summaries as Centroids for Interpretable and Scalable Text Clustering},
  author    = {Diaz-Rodriguez, Jairo},
  booktitle = {ICLR 2026},
  year      = {2026}
}
